\documentstyle[aps,preprint,epsf]{revtex}
\begin{document}
\draft
\title{ Weak Localization in  the Conductance Peaks of
  Coulomb Blockade Quantum Dots}

\author{ Y. Alhassid}

\address{ Center for Theoretical Physics, Sloane Physics Laboratory,
	 Yale University, New Haven, Connecticut 06520, USA}
           	
\maketitle
\begin{abstract}
 We derive closed expressions  for the universal  weak localization peak 
 of the average conductance peak heights in Coulomb blockade quantum dots
 in the crossover from orthogonal to unitary symmetry.  The  scale for the
 crossover is independent of the number of channels in each lead, in contrast 
with the case of open dots.   The functional form of the weak localization peak 
is independent of temperature.  We also derive analytically the  variance of 
the conductance peak heights  as a function of the crossover parameter. 
 
\end{abstract}


\narrowtext

Quantum dots  are formed in the interface between
 layers of semiconductors by the electrostatic confinement of electrons
 in a two-dimensional electron gas. 
 The transport properties (conductance) of such quantum dots can be measured 
by connecting them to external leads.
In ballistic dots, the transport is dominated by scattering from the boundaries.
Since dots with several hundred electrons are usually characterized
 by irregular shape, the scattering  from the boundaries results in
 chaotic motion of the electrons.
 The conductance in such dots displays universal statistical  
fluctuations that are a signature of quantum chaos and that can be
 described by random matrix theory (RMT) \cite{Me95}.  
 Earlier studies of quantum dots focused mostly on open dots,
 where the movement of the electron into the dot is allowed classically 
\cite{MW93}. 
As the point contacts are pinched off, the dot becomes
closed, i.e., effective tunnel barriers are formed in the interface between the
dot and the leads.  In such dots  the conductance is dominated by resonant
 tunneling through electron resonances that are narrow compared with their 
average spacing. This leads  to a series of narrow peaks in the conductance
 as a function of  gate voltage,  known as Coulomb blockade peaks.
A statistical theory based on RMT  for
 the fluctuations of the conductance peak heights 
 was developed in Ref. \cite{JSA92} and  was recently confirmed
 experimentally \cite{Chang96,Marcus96}.   

  The statistics of the peak heights are universal but sensitive to the
 underlying symmetries.  For conserved time-reversal symmetry 
 (no magnetic field) the appropriate  ensemble  is the Gaussian 
orthogonal ensemble (GOE), while for broken time-reversal symmetry
 the statistics are described by the Gaussian unitary ensemble (GUE).  
 The crossover
from preserved to fully broken  time-reversal symmetry is also universal. 
The distribution of the wavefunction intensity at a fixed spatial point   was
 calculated by the supersymmetry method in Ref. \cite{FE94}.  However 
 the fluctuations of  the conductance in the crossover regime  
depend on  the wavefunction fluctuations at several  spatial points. 
 This statistics and  the corresponding
distributions of the peak heights were  derived as a function 
of a dimensionless crossover parameter  in Ref. \cite{AHW96}.  
This crossover parameter is linear in the magnetic field
 that induces the  symmetry breaking.   

 One of the interesting phenomena  associated with the 
onset of broken time-reversal symmetry in open dots is the
 suppression of weak localization.   Weak localization in the absence of 
magnetic field originates from the constructive interference of  pairs of
 time-reversed 
classical trajectories, which enhances  the return probability. This results
 in a smaller average conductance. When a magnetic field is applied,
 the  constructive interference  is destroyed and the weak
 localization is suppressed.
 In open dots,  the semiclassical theory yields a  Lorentzian shape for  the
 weak localization peak  \cite{BJS93}.  An exact expression
  for open dots was derived from RMT using the supersymmetry 
method \cite{PW94}, and approaches a Lorentzian for a large number
 of open channels.

  Weak localization also occurs in almost closed dots, 
where the average 
conductance peak height was found to be smaller  in the GOE case than in
 the GUE case \cite{JSA92}.  On the other hand, no weak localization effect
 is found in the average of the conductance minima \cite{AG96,Marcus97}.
 In this paper we derive closed  expressions  for the  weak localization peak
of the conductance maxima in closed dots as a function of the crossover 
parameter and for any number 
of equivalent channels in the leads.  We also study the temperature 
dependence of the weak localization peak and find that, 
when measured in units of the average GUE conductance,
the peak is  temperature-independent
  (assuming no phase breaking).  We find that the width
 of the peak is only weakly dependent  on the number of channels in the
 leads, in contrast to the case of open dots, where the width of the peak
 is proportional to the square  root of the number of open channels.   We also
  derive closed expressions for  the variance of the conductance peak  as a
 function of the crossover parameter.

 For  almost closed dots, the resonances in the dots are isolated
 and their  average width is small compared
with the mean level spacing $\Delta$. 
At  temperatures that are low compared with the mean level 
spacing  ($kT \ll \Delta$), only one resonance contributes to 
a given Coulomb blockade peak, and  the conductance peak amplitude $G$ 
 is  $G =\frac{e^2}{h}\, \frac{\pi \bar{\Gamma}}{4 kT} g$
where \cite{Be91}
\begin{eqnarray}\label{Peak}
   g=  {1 \over \bar{\Gamma}} \frac{\Gamma^l \Gamma^r}{\Gamma^l
 + \Gamma^r} \;.
\end{eqnarray}
Here $\Gamma^{l(r)}$  is the width
of a resonance to decay into  the
 left (right) lead and $\bar{\Gamma}$ is the average width per
 channel. 
Note that the quantity $g$ is dimensionless and  temperature-independent.
 In general we assume that the left (right) lead has  $\Lambda^{l(r)}$ open 
channels such that $\Gamma^{l(r)} = \sum_c |\gamma_{c}^{l(r)}|^2$, where
$\gamma_{c}^{l(r)}$ is the partial amplitude to decay into channel
$c$ on the left (right).  
 In $R$-matrix theory,  the partial  width amplitude  $\gamma_c$
can be expressed  as a scalar product (defined over the dot-lead interface) 
of the resonance wavefunction $\psi$ and the channel wavefunction $\phi_c$
through
$\gamma_{c} = \langle \mbox{\boldmath$\phi$}_c
| \mbox{\boldmath$\psi$} \rangle \propto  
 \int \! dS\, \phi_c^\ast(\bbox{r}) \psi (\bbox{r})$.
We assume for simplicity  that
 the $\Lambda$ channels are uncorrelated and equivalent, i.e.   
 $\overline{\gamma_c^\ast \gamma_{c^\prime}} = 
\bar{\Gamma} \delta_{c c^\prime}$. All average partial widths 
are assumed to be independent of the magnetic field. We also assume that
 the dot has symmetric leads, so that the average partial widths are 
the same in the left and right leads.

 For a ballistic dot with chaotic dynamics of the electrons, we model  the  
Hamiltonian $H$ of the dot by an ensemble of random matrices.  When
 we apply a magnetic field to the dot,  the ensemble describing the
crossover from  GOE to GUE  is \cite{Mehta91,Bohigas91}

\begin{eqnarray}\label{TranEns}
  H=S+i\alpha A  \;,
\end{eqnarray}
where $S$ and $A$ are uncorrelated symmetric and antisymmetric real
 matrices, respectively,  chosen from Gaussian ensembles of the same variance,
and $\alpha$ is a real parameter. 
The transition parameter  describing  the crossover  is given by the 
ratio of the root-mean-square (rms) of a typical 
symmetry-breaking matrix element
 to the mean-level spacing \cite{French82}
$\lambda = (\overline{H^2}_{break})^{1/2} / \Delta = \alpha \sqrt{N} / \pi$.
 Alternatively  $2 \pi \lambda = \sqrt{\tau_H/\tau_{mix}}$,
 where $\tau_H = h/\Delta$ is
 the Heisenberg time and $\tau_{mix}$ is the mixing time defined
 by the spreading width  $\hbar/\tau_{mix} = 2 \pi  
\overline{H^2}_{break}/\Delta$
 due to the interaction that breaks time-reversal symmetry \cite{PW94}.
The spectral properties of the ensemble (\ref{TranEns}) make the
 complete crossover  for $\lambda  \sim1$. 
 The average dimensionless conductance
 peak  height $\bar{g}(\lambda)$ can be calculated by
 averaging (\ref{Peak}) over the ensemble  (\ref{TranEns}). 

  The statistics of the partial width amplitudes $\gamma$ were derived
  in Ref. \cite{AHW96}.  The  components
 $\psi_\mu$ of the resonance wavefunction (in a fixed basis)
  are decomposed into their  real and imaginary parts  
$\psi_\mu = \psi_{\mu R} +  i \psi_{\mu I}$ in a principal frame in the 
complex plane (determined by $\sum_{\mu=1}^N \psi_{\mu R} \psi_{\mu I} = 0$),
 and  a parameter $t$ is defined by  $t^2  \equiv  \sum_\mu 
\psi_{\mu I}^2  /  \sum_\mu \psi_{\mu R}^2$.
   At a {\it fixed} value of $t$, the real and imaginary parts of  a finite 
number $\Lambda$  of
 components ($\Lambda \ll N$) are independent Gaussian variables.  Except in 
the GOE and GUE limits (where $t=0$ and $t=1$, respectively),  $t$ is
 not sharp but  fluctuates according to a known distribution 
\cite{FE94,AHW96,LBB97} 
\begin{eqnarray}\label{tDist}
 P_\lambda(t) = \pi^2 { 1-t^4 \over t^3} \lambda^2
 e^{-{\pi^2 \over 2} \lambda^2 \left(t- 1/t \right)^2}
 \left\{ \phi_1(\lambda) + \left[ \frac{1}{4} \left( t + \frac{1}{t} \right)^2
 - {1 \over 2\pi^2 \lambda^2 } \right]  \left[ 1 - \phi_1(\lambda) \right] 
\right\} \;,
\end{eqnarray}
where $\phi_1(\lambda)= \int\limits_0^1 e^{ - 2 \pi^2 \lambda^2 (1 - y^2)} dy$.
These fluctuations of $t$ are responsible for the long-range correlations of 
wavefunction intensities in the crossover from GOE to GUE \cite{FE96}.

 Applying a similar decomposition for the partial width amplitudes 
 $\gamma_c = \gamma_{cR} + i \gamma_{cI} = 
\langle \bbox{\phi}_c | \bbox{\psi}_R \rangle +
 i  \langle \bbox{\phi}_c | \bbox{\psi}_I \rangle$ in the
 principal frame of $\bbox{\psi}$, it is
 found that at fixed $t$,  $\bbox{\gamma}_R$ and $\bbox{\gamma}_I$ 
are Gaussian statistically  independent variables  with 
 $\overline{\gamma^2_{cI}} = t^2  \overline{\gamma^2_{cR}}$.
 The joint distribution of the partial widths  is obtained by averaging
 the  conditional distribution $P (\bbox{\gamma} | t)$
 over the   $t$ distribution,
$P_\lambda(\mbox{\boldmath$\gamma$}) = \int\limits_0^1 P_\lambda(t) 
 P(\mbox{\boldmath$\gamma$} |t) dt$.  
 
    Using Eq.  (\ref{Peak}) and the statistical independence of the
 partial widths in the left and right leads at {\it fixed} $t$  (i.e., 
$P(\Gamma^l, \Gamma^r|t)= P(\Gamma^l | t) P(\Gamma^r | t)$ where
$P(\ldots |t)$ are conditional probabilities),  the
 average conductance  peak height can be  written as 
\begin{mathletters}
\label{average}
\begin{eqnarray}
 \bar{g}(\lambda) & =  & \left\langle  \int_0^\infty d\Gamma  
Q^{(1)}(\Gamma,t) / \Gamma \right\rangle  \label{averagea} \\
Q^{(1)}(\Gamma,t) & \equiv &
 \int_0^\Gamma d\Gamma^l  \Gamma^l P(\Gamma^l |t)
 (\Gamma - \Gamma^l)P(\Gamma - \Gamma^l | t) \;, \label{averageb}
\end{eqnarray}
\end{mathletters}
where $\langle \ldots \rangle$ denotes here and in the following an
 average over the $t$ distribution $P_\lambda(t)$ given by Eq. (\ref{tDist}).  
 The function $Q^{(1)}(\Gamma,t)$ 
is  the convolution of $\Gamma^l P(\Gamma^l)$ with itself,  and its Laplace
 transform is   $Q^{(1)}(s,t) = (\partial P/\partial s)^2$ where 
 $P(s,t) \equiv \int_0^\infty d \Gamma e^{-s \Gamma} P(\Gamma |t)$ is
the Laplace transform of the width distribution.
$P(s,t)$  is easily calculated using the Gaussian nature of the partial width 
amplitudes at fixed  $t$ and $\Gamma = \sum_c\left( |\gamma_{cR}|^2 +
 |\gamma_{cI}|^2 \right)$.  
Assuming one-channel symmetric leads (i.e.
 $\overline{\Gamma^l} = \overline{\Gamma^l} =\overline{\Gamma}$)
 and measuring  $\Gamma$ in units of $\overline{\Gamma}$, 
 we find  $P_1(s,t) = \left( 1 +s^2/x^2 + 
2s \right)^{-1/2}$ where $ x  \equiv (t^{-1} + t)/2$, and thus
 $Q^{(1)}_1(s,x) =  x^2 (s+x^2)^2 (s^2 + 2sx + x^2)^{-3}$. 
The inverse Laplace transform
$Q^{(1)}_1(\Gamma,t)$ is calculated by the residue theorem
 using the poles (of degree three) of $Q_1(s,x)$ at  $s = t$ and $s=1/t$.  
 After  performing the integral in Eq.  (\ref{averagea}), we  find
  the average conductance as a function of $\lambda$  to be
\begin{eqnarray}\label{wlocal1}
\bar{g}(\lambda) = \frac{1}{4} +  \left\langle 
 \left(\frac{t}{1-t^2}\right)^2 \left({2 t^2 \over 1- t^4} \ln t  +\frac{1}{2}
 \right)
\right\rangle \;.
\end{eqnarray}
 The GOE and GUE limits are obtained for $t \to 0$ and $t \to 1$, 
respectively, where $\bar{g}_{\rm GOE}=1/4$ and 
$\bar{g}_{\rm GUE}=1/3$ \cite{JSA92}.

 The above method can be generalized for any number  $\Lambda$  of
equivalent and uncorrelated channels (in each of the symmetric leads). 
Measuring $\Gamma$ in units of the average 
 resonance width per channel,  we find $P_\Lambda (s,t)  = 
\left(1 +s^2/ x^2
+ 2s \right)^{-\Lambda/2}$ and 
$Q^{(1)}_\Lambda(s,t) = \Lambda^2  x^{2 \Lambda} 
(s +  x^2)^2 (s^2 + 2 x^2 s +  x^2 )^{-\Lambda -2}$. 
Using (\ref{average}) we find that  $\bar{g}_\Lambda$ changes from 
$\Lambda^2/ 2(\Lambda+1)$  in the GOE limit    ($x \to \infty$)
 to $\Lambda^2/ (2\Lambda+1)$ in the GUE limit ($x \to 1$).
In  the crossover regime we find  that  $\bar{g}_\Lambda(\lambda)$ has the form 
\begin{eqnarray}\label{wlocalL}
\bar{g}_\Lambda (\lambda) = \frac{\Lambda^2}{2(\Lambda+1)} +
 (-)^{\Lambda+1} \Lambda \left( \begin{array}{c} 2 \Lambda \\ \Lambda-1 
\end{array} \right)  \left\langle 
 \left(\frac{t}{1-t^2}\right)^{2 \Lambda} \left[{2 t^2 \over 1- t^4} 
\ln t  - \sum_{n=0}^{\Lambda-1} a_n \left( {1-t^2 \over t} \right)^{2n} \right]
\right\rangle \;.
\end{eqnarray}

  The coefficients $a_n$ in Eq. (\ref{wlocalL}) can be determined from
 the known GUE limit ($t \to 1$) of $\bar{g}_\Lambda$. 
For  the limit on the r.h.s. of  (\ref{wlocalL}) to exist when $t \to 1$,
 the coefficients
  $a_n$ for $n=0,1,\ldots,\Lambda-1$  must be equal to the corresponding 
expansion coefficients  of the function $2t^2/(1 - t^4) \ln t$ in 
 powers of $\left[(1-t^2)/t\right]^2$, and are therefore independent of 
$\Lambda$. The value of $\bar{g}_\Lambda$ in the GUE limit also determines the 
coefficient $a_\Lambda$ 
 to be $a_\Lambda = (-)^{\Lambda +1}( \Lambda !)^2/ 2((2 \Lambda+1)!$. 
Since the coefficients $a_n$ are independent of $\Lambda$, we conclude 
\begin{eqnarray}
 a_n = (-)^{n+1} { (n!)^2 \over 2(2n+1)!} \;.
\end{eqnarray}

\begin{figure}

\vspace{3 mm}

\epsfxsize=12 cm
\centerline{\epsffile{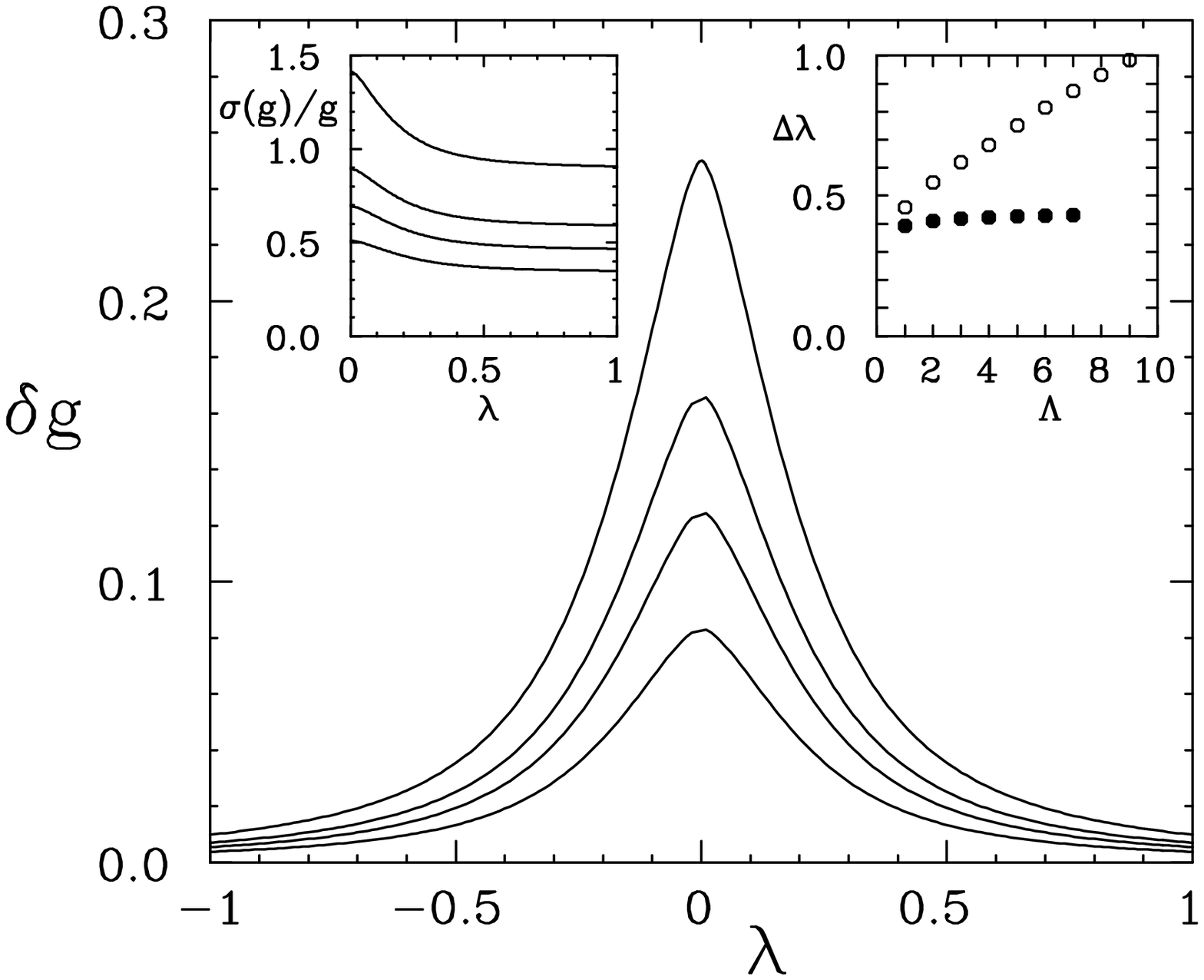}}

\vspace{1 cm}

\caption
{ The weak localization peak in closed dots (measured in units of the average
 GUE conductance) $\delta g(\lambda)/\bar{g}^{\rm GUE}$  versus the crossover
 parameter $\lambda$. Shown are the cases with
 $\Lambda=1,2,3$ and $5$ channels (peak is decreasing as $\Lambda$ increases). 
Right inset: The FWHM of the weak localization peak $\Delta \lambda$ as a 
function of the number of channels $\Lambda$ in each lead for closed and open dots (solid 
and open circles, respectively).  
Left inset:  The  ratio  $\sigma(g) /\bar{g}$ between  the rms and average 
values of the 
conductance peak as a function of the crossover parameter $\lambda$ for  
Coulomb blockade dots
with $\Lambda=1,2,3$ and $5$ channels ($\sigma(g)/\bar{g}$ is 
smaller for larger $\Lambda$). 
}
\label{fig1}

\end{figure}

The function $\delta g_\Lambda(\lambda) \equiv \bar{g}^{\rm GUE}_\Lambda - 
\bar{g}_\Lambda(\lambda)$,
which  is peaked at $\lambda=0$ (GOE limit) and approaches zero for
 $\lambda \to \pm \infty$, is known as the weak localization peak. 
Its analytic form is plotted in Fig. \ref{fig1} (in units of the average
 GUE conductance peak) for $\Lambda=1, 2, 3$ and $5$.
The size of the peak is   $\delta g_\Lambda(0) = \bar{g}^{\rm GUE}_\Lambda - 
\bar{g}^{\rm GOE}_\Lambda = \Lambda^2/[2(\Lambda+1)(2\Lambda+1)]$.  
For open dots with $\Lambda$ equivalent channels (with transmission
 coefficient 1) in each lead the size of the weak localization peak is 
$\Lambda/[2(2\Lambda+1)]$ \cite{PW94},   
so that the weak localization peak in closed dots  (measured in units 
of $\frac{2 e^2}{h}\, \frac{\pi \bar{\Gamma}}{4 kT}$)
  is smaller by a factor of  $\Lambda /(\Lambda +1)$ than the one in open dots
 (measured in units of $e^2/h$).
In units of  the average GUE conductance height, 
we find   $\delta g_\Lambda(0)/\bar{g}^{\rm GUE}_\Lambda 
= 1/[2(\Lambda+1)]$ for closed dots, compared with $1/(2 \Lambda +1)$ for open dots.

   It is interesting to compare  the dependence
 of the width of the weak
 localization peak on the number of channels in open versus closed dots.
The right inset of  Fig. \ref{fig1} shows the full width
 at half maximum (FWHM) of the weak
 localization peak as a function of the number of channels for both closed 
(solid circles) and open (open circles) dots.  The FWHM in closed dots is 
approximately independent of the number of channels, unlike the case of 
open dots where the FWHM  increases as $\sqrt{\Lambda}$
 (for large $\Lambda$). In open dots,
 the crossover scale in the conductance is determined  by the competition 
between the mixing time $\hbar/\tau_{mix} = 2\pi \Delta \lambda^2$ 
(associated with the breaking of time-reversal symmetry) and the decay time of 
resonances in the dot $\hbar /\tau_{dec} = (\Delta /2\pi) \Lambda$.
  The crossover in the conductance occurs when
 $\tau_{dec}/\tau_{mix} = 4 \pi^2 \lambda^2/\Lambda \sim1$
 i.e., for $\lambda^{open}_{cr} \sim \sqrt{\Lambda} /2\pi$.   The decay time
 in an open dot
 is shorter than the Heisenberg time by a factor of $\Lambda$, and for one open
 channel these two times are equal. For closed dots on the other hand, the 
decay time 
is much longer than the Heisenberg time (since the average level width is
 much smaller than $\Delta$), irrespective of the number of channels. In such
 a case the relevant time scale that competes with the mixing time is the 
Heisenberg time, and the crossover  occurs on a scale of 
$\lambda^{closed}_{cr} \sim 1$,
 independent of the number of channels and comparable to  the value
 of $\lambda_{cr}$ for  an open dot with $\Lambda=1$. This is confirmed
 by the exact RMT results
 of Fig. \ref{fig1}, where the full width at half maximum (FWHM) of the
 weak localization peak  is $\Delta \lambda\approx 0.4$, irrespective of the
 number of channels in each lead.  

If time-reversal symmetry is broken by a magnetic field, then 
$\lambda=\Phi/\Phi_{cr}$,
where $\Phi_{cr}$ is the crossover flux  (defined as  the flux 
where $\lambda=1$). It is possible to estimate $\Phi_{cr}$ semiclassically 
assuming single-particle chaotic motion \cite{JSA92,BG95}.  The rms
 of  the electron action 
difference between a pair of time-reverse orbits is given by
$[(\overline{\Theta^2})^{1/2}/{\cal A}](\Phi/\Phi_0)$ (in units of $\hbar$), where 
$(\overline{\Theta^2})^{1/2}$ is the rms area accumulated by the electron 
and ${\cal A}$ is the area of the dot. Since  
the area accumulation in a chaotic dot is diffusive,
we have for an open dot  $(\overline{\Theta^2})^{1/2}/{\cal A} = 
{\alpha_g}^{1/2} \sqrt{\tau_{cr}/\tau_{dec}}$, where $\tau_{cr}$ is the time 
scale for the electron
 to cross the dot, and  $\alpha_g$ is a geometrical factor that is
 dot specific \cite{BG95}. 
For the breaking of time reversal symmetry, we require the rms action difference
 to be of order $1$, leading to 
$\Phi^{open}_{cr}/\Phi_0 = {\pi \alpha_g}^{-1/2} \sqrt{\Lambda} 
\sqrt{\tau_{cr}/\tau_H}$. 
For the closed dot, on the other hand, $\tau_H$ replaces 
$\tau_{dec}$ to give a $\Lambda$-independent crossover flux 
of  $\Phi^{closed}_{cr}/\Phi_0 = {\pi \alpha_g}^{-1/2} 
\sqrt{\tau_{cr}/\tau_H}$.  
 A weak localization  of the average 
 conductance peak height was observed experimentally  in semi-open dots 
\cite{Marcus97} with an FWHM of $\approx 6.2$ mT,  
 corresponding to $B_{cr} \approx 14.9$ mT or $\Phi_{cr}  \approx 2 \Phi_0$.
The desymmetrized stadium billiard  in a uniform magnetic field 
gives $\Phi_{cr} \approx \frac{2}{3} \Phi_0$ \cite{BG95} with about the
 same number of electrons ($\sim 1000$). Thus the experimental result is
 about three times larger than the billiard model estimate, similar to the
 discrepancy estimated for  the correlation field \cite{AHW96}. 
 The exact discrepancy is uncertain  because
 of the unknown geometrical
 factor of the dot used in the experiment.  The discrepancy does not indicate 
a breaking of RMT (which only predicts the universal form of the 
correlator \cite{AA96}
and not its magnetic flux  scale), but  rather that effects beyond 
single-particle 
 dynamics are important.  Indeed,  a model that takes into account  
electron-electron interactions in the dot 
gives a  correlator whose form is consistent with the RMT correlator, but  with
a correlation field  that is larger by about a factor of three as
 compared with a non-interacting model \cite{BS98}.

  It is also possible to calculate in closed form the variance of the
 conductance as a function of  $\lambda$. 
 The second moment of the conductance is given by  
 $\overline{g^2}(\lambda) = \left\langle  \int_0^\infty d\Gamma  
Q^{(2)}(\Gamma,t) / \Gamma^2 \right\rangle$, where $Q^{(2)}$ is the convolution
 of  $({\Gamma^l})^2 P(\Gamma^l | t)$ with itself. The Laplace transform
 is then calculated from  $Q^{(2)}(s,t) = \left[d^2P(s | t)/ds^2 \right]^2$.   
For one-channel leads we find 
\begin{eqnarray}
\overline{g^2}(\lambda)  = {3 \over 16} + {27 \over 2} 
\left\langle \left( {t \over 1+t^2} \right)^2  \left( {t \over 1-t^2} 
\right)^4 \left[ {1+t^2 \over 1-t^2} \ln t +1 + \frac{1}{12}
 \left( { 1-t^2 \over t}\right)^2 -\frac{2}{27} 
\left( { 1-t^2 \over t}\right)^4 \right]
 \right\rangle \;.
\end{eqnarray}
$\overline{g^2}$ increases from $3/16$ (GOE) to $1/5$ (GUE), while the variance
 $\sigma^2(g)=\overline{g^2} -\bar{g}^2$ decreases from $1/8$ to $4/45$.
Closed expressions for
$\overline{g^2_\Lambda}(\lambda)$ can be similarly derived  for any
 number of equivalent channels $\Lambda$. 
 The left inset of Fig. \ref{fig1} shows the analytically calculated
  $\sigma(g) /\bar{g}$ as a function of $\lambda$ for $\Lambda=1,2,3$ 
and $5$ channels.  We see that the distribution becomes sharper in
 the crossover from conserved to broken time-reversal symmetry.
  In the GOE and GUE limits we have
\begin{eqnarray}
\sigma(g) /\bar{g} =
 \left\{ \begin{array}{cc} \left[{(\Lambda^2 + 5\Lambda +2) \over \Lambda^2
 (\Lambda+3)}\right]^{1/2}  & ({\rm GOE})
\\    \left[{(2\Lambda^2 +
 5\Lambda +1) \over 2\Lambda^2 (2\Lambda+3)}\right]^{1/2}
& ({\rm GUE})
\end{array}  \right. \;.
\end{eqnarray}
For large $\Lambda$, this quantity decreases in the crossover 
 from $1/\sqrt{\Lambda}$ (GOE)  to $1/\sqrt{2\Lambda}$ (GUE), i.e.,
 effectively doubling the number of channels.

  At finite temperature, several levels $i$ contribute to the 
conductance peak with known thermal weights
 $w_i(T)$ \cite{Be91,AGS97} which are determined by the single-particle  
 spectrum and the number of electrons in the dot. 
 The dimensionless peak height is given by  
$g = \sum_\i w_i (T) g_i$, where  $  g_i$ 
are the level conductances as in Eq. (\ref{Peak}), but 
with $\Gamma^{l,r}$ depending now on the level $i$. 
In  calculating the statistics of the peak heights at finite $T$, 
a good approximation
 is to neglect the fluctuations in the dot's spectrum, and $w_i(T)$ are 
then fixed. 
 Since $\bar{g}_i(\lambda)$ are independent of $i$, the weak localization
 peak factorizes $\delta g(\lambda,T) \approx \left[\sum_i w_i(T) \right]
 \delta g (\lambda)$, where $\delta g(\lambda)$ is the weak localization
 peak at temperatures much smaller than $\Delta$. In particular, 
since $\bar{g}^{\rm GUE} (T)
  \approx \left[\sum_i w_i(T) \right] \bar{g}^{\rm GUE}$,
 the {\it scaled}  weak localization peak  $\delta g/ \bar{g}^{\rm GUE}$ is
 temperature-independent (assuming no phase breaking).
 
Another quantity of interest is the size of the  weak localization peak 
relative to the rms  of the (GUE) peak-heights distribution.  For dots with 
 one-channel leads, we find 
 $\delta g(0)  / \sigma^{\rm GUE}(g)
 \approx(\sqrt{5}/ 8) \left[ (\sum w_i)^2 / \sum w_i^2 \right]^{1/2}$, 
which  increases with temperature. At  high temperatures, 
the number of levels that contribute to a given peak height  is of the order
 of $T/\Delta$, and  ${\delta g(0) / \sigma_{\rm GUE}(g)}
 \propto (T/\Delta)^{1/2}$.
At low temperatures it is difficult to  measure  the weak localization peak 
 because its size  is comparable
 to the conductance  fluctuations.  In open dots, the large number
 of open channels enhances the peak relative to the fluctuations by
 $\sqrt{\Lambda}$.  In closed dots (with $\Lambda=1$), the temperature
 can be used instead to enhance 
 the peak relative to the  background fluctuations.

In conclusion, we have derived  in closed form 
the weak localization peak of Coulomb
blockade peak heights in quantum dots. In the absence of
 phase breaking,  this peak is temperature-independent up to an 
overall scale.

We acknowledge C.M. Marcus for helpful discussions.
  This research was supported in part by  DOE Grant
DE-FG02-91ER40608.

\end{document}